\documentclass[prd,aps,
tightenlines,
superscriptaddress,showpacs,nofootinbib,eqsecnum,amsfonts,amsmath,epsf
]{revtex4}
\usepackage{bm,graphics,graphicx,epsfig,color
}

 \def\gsim{\mathrel{
 \rlap{\raise 0.511ex \hbox{$>$}}{\lower 0.511ex
 \hbox{$\sim$}}}}
 \def\lsim{\mathrel{
 \rlap{\raise 0.511ex \hbox{$<$}}{\lower 0.511ex
 \hbox{$\sim$}}}}
\begin{document}
\title{ Parameter estimation of coalescing supermassive  black
hole binaries with LISA}

\author{K.G.\ Arun} \email{arun@rri.res.in} \affiliation{Raman Research
Institute, Bangalore 560 080, India}
\begin{abstract}
Laser Interferometer Space Antenna (LISA) will routinely observe
coalescences of supermassive black hole (BH) binaries
up to very high redshifts.
LISA can  measure mass parameters of such
coalescences to a relative accuracy of $10^{-4}-10^{-6}$, for sources at a
distance of 3 Gpc.
The problem of parameter estimation of massive nonspinning binary black holes
using post-Newtonian (PN) phasing formula is studied in the context of
LISA.  Specifically, the
performance of the 3.5PN templates is contrasted against its
2PN counterpart using a waveform which is averaged over the
LISA pattern functions. The improvement due to the higher
order corrections to the phasing formula is examined by calculating 
the errors in the estimation of mass parameters at each order.
The estimation of 
the mass parameters ${\cal M}$ and $\eta$ are significantly enhanced by using
the 3.5PN waveform instead of the 2PN one.
For an equal mass binary of
$2\times10^6M_\odot$ at a
luminosity distance of 3 Gpc, the improvement
 in chirp mass is $\sim 11\%$ and that of $\eta$ is $\sim 39\%$.
Estimation of coalescence time $t_c$ worsens by $43\%$. 
The improvement is larger for the unequal mass binary mergers.
These results are compared to the ones obtained using
a non-pattern averaged waveform. The errors depend very much on the
location and orientation of the source and general conclusions cannot be
drawn without performing  Monte Carlo simulations.
Finally the effect of the choice of the lower frequency cut-off for LISA on the parameter estimation is studied.

\end{abstract}
\date{\today}
\widetext
\maketitle
\section{Introduction}\label{sec:intro}
Most of our near-by galaxies harbour supermassive black holes (BH) at their centre~\cite{Richstone98}. If this is  the case, then merger of such galaxies
would produce a binary system composed of two supermassive BHs.
Simulations indicate that there could be primordial supermassive BH
binaries at the centres of the first galaxies~\cite{BrommLoeb03}.
 An understanding of the formation and evolution of these binaries is
very important from the view point of cosmology and structure formation.
Many of  these binaries coalesce under gravitational wave (GW) radiation
reaction within Hubble time.
Supermassive BH binaries in the mass range $10^4$--$10^7 M_{\odot}$  will
emit gravitational waves (GW) of frequency $10^{-4}$--$10^{-1}$ Hz during its
adiabatic inspiral phase which can be observed by the proposed
space-borne GW missions such as LISA~\cite{Bender95,Danzmann97}
 with high ( $\sim$ a few thousands) signal to noise ratio (SNR)
  up to  very high redshifts ($\sim 10$).

  Different authors have investigated the
implications of these observations in the context of astrophysics,
cosmology and testing general relativity and its alternatives.  LISA observations of BH coalescences can
be used to study the growth of BHs as the universe evolved and for mapping the
distribution of BHs as function of the
redshift~\cite{Hughes02,BBW05a,BBW05b}. LISA will be able to
measure luminosity distances to the sources with an accuracy $\sim 1-10\%$.
If the redshift associated with the event is known by electromagnetic
observations, these sources can be used as very high precision standard
candles and to study the distance-redshift
relation~\cite{Schutz86,HolzHugh05}. Ref~\cite{Miller05} discussed the potential
of LISA to observe binaries containing a BH in the intermediate mass
regime ($\sim 10^3 M_\odot$) and use it as a probe of strong field aspects of
gravity.

LISA could probe many strong gravitational field effects which are
not possible to explore with other observational means.
Both inspiral and ring-down GW signals can be used for this.
Refs~\cite{BHspect04,BCW05} studied the possibility of using the
quasi-normal mode oscillations to test the no-hair theorem of general
relativity since these modes will be characterized {\it only} by the
mass and angular momentum of the BH (in general relativity). Hughes and Menou examined another
possibility~\cite{HughMen05} if LISA detects both inspiral and ring-down signals
from the same source. By measuring the total mass independently
 from both the signals
one can estimate the mass difference which will be the mass-energy lost
due to GWs. They suggested that an extension of this idea including
spin effects could in principle test the BH area theorem.
Further, inspiral of a stellar mass BH into a SMBH will be another
interesting source for LISA using which many properties of the central
SMBH can
be probed including the possibility to map the spacetime following the
geodesics of the stellar mass BH~(See e.g.\cite{Ryan97,CollHugh04}) and
measuring the multipole moments of the spacetime.

LISA will provide an unique opportunity to test general relativity and
its alternatives. Will and his co-workers have discussed
the potential of LISA to test
general relativity as well as its alternatives like Brans-Dicke theory
and massive graviton theories~\cite{Will98,WillYunes04}.
Recently this issue was discussed in a more realistic
scenario of spinning binaries using 2PN phasing~\cite{BBW05a,BBW05b}.
Blanchet and Sathyaprakash proposed another test based
on post-Newtonian (PN) GW phasing formula by  measuring the 1.5PN GW
tail effect and showing how it can be used as a test of general relativity~\cite{BSat94,BSat95}. This proposal was recently
generalized to higher order terms in the phasing formula by Arun {\it et
al}~\cite{AIQS06a} and it was argued that such a test would allow one to probe the nonlinear
structure of gravity~\cite{AIQS06b}.

A very accurate parameter extraction scheme is central to performing all
these analyses.
A parameter
estimation scheme based on matched filtering, similar to that for 
 the ground based detectors such as LIGO and VIRGO, will be
employed for LISA also. An efficient matched filtering would in turn
demand a very accurate model
of the gravitational waveform. In order to compute the gravitational
waveform from a compact binary system, one  solves the two-body
problem in general relativity perturbatively using different approximation
schemes since no exact solutions for this problem exist till date.
The final
waveform can be expressed as a post-Newtonian expansion which is a 
power series in $v/c$ where $v$ is the
gauge independent velocity parameter characterising the source (See
Ref.~\cite{Bliving} for an exhaustive review on the formalism).
In our notation ${v\over c}$ refers to half a PN order.

Since the information about the phase is more important for the process of matched filtering, one uses  a simplified model of the inspiral waveform (the so called restricted waveform)
where phase is modelled to a high PN order retaining the Newtonian amplitude.
In doing so, one is neglecting the effect of other
harmonics~\cite{BIWW96,ABIQ04} in the
amplitude of the wave and also the higher order PN corrections to the
 dominant harmonic
at twice the orbital frequency. In the present study,
we deal only with the restricted waveform in the Fourier domain obtained using
stationary phase approximation.
The phasing formula for nonspinning binaries, is presently complete up to
3.5PN order~\cite{BDIWW95,B96,BFIJ02,BDEI04}.

The implications of the higher PN order phasing in the context of parameter
estimation problem has been investigated by different authors.
Based on the framework set up by Refs~\cite{Finn92,FCh93}, Cutler
and Flanagan~\cite{CF94} investigated the importance of the 1.5PN phasing
formula~\cite{3mn}. The effect of including spin-orbit coupling parameter
at 1.5PN order
into the space of parameters was one of interesting issues addressed.
Two independent works by Kr{\'o}lak {\it et al.}~\cite{KKS95} and 
Poisson and Will~\cite{PW95} analysed the problem of parameter
estimation using the 2PN phasing formula of Ref.~\cite{BDIWW95}. Inclusion
of the spin-spin coupling term at 2PN and its effect on the errors of
other parameters was the focus of their analysis.

Recently Arun {\it et al.}~\cite{AISS05} investigated 
the effect of the 2.5, 3 and 3.5PN terms for the parameter estimation of
nonspinning binaries (a similar work was carried out independently by Berti and
Buonanno~\cite{BB04}).
Using covariance matrix calculations, they inferred that by employing
the 3.5PN phasing instead of the
2PN one, the improvement in estimation of errors in chirp mass and symmetric
mass ratio can be as high as $19\%$ and $52\%$ respectively for the ground based
detectors such as LIGO and VIRGO.

 Cutler was one of the first to address the problem of
parameter estimation in the LISA context~\cite{Cutler98}. He used
the 1.5PN waveform including the spin-orbit effect and studied the
estimation of errors associated with the mass parameters as well as distance
and angular resolution of the binary. Seto investigated the effect of
finite arm length of LISA using 1.5PN phasing~\cite{Seto02}. Vecchio revisited the parameter
estimation with the 1.5PN waveform~\cite{Vecchio04} where he used the waveform
for circular orbit but with ``simple precession" (as opposed to the 
non-precessing
case of ~\cite{Cutler98}) and examined the
implications of it for the estimation of distance and angular
resolution.  Various aspects of the 2PN
parameter estimation, such as the spin-spin coupling, was investigated
by different authors~\cite{Hughes02,BBW05a,BBW05b}. Ref~\cite{BBW05a}
studied the effect of spin terms in testing alternate theories of
gravity with the LISA observations.
Refs~\cite{Hughes02,BBW05b}  also
addressed the issue of mapping the merger history of massive BHs
using LISA observations in the 2PN context. While all these calculations are within the
restricted waveform approximation where the PN corrections to
the amplitude is completely neglected, there have been investigations
about the effect of including these amplitude corrections in the context
of parameter estimation~\cite{MH02,HM03,SinVecc00a,SinVecc00b}.

Other than the covariance matrix approach, which is
valid only in the high SNR limit,  there have
been proposals in literature addressing the parameter estimation problem
using Monte-Carlo methods. In Ref~\cite{BalSatDhu96}, the authors  compared
the error estimates obtained using the covariance matrix with
the  Monte-Carlo simulations. Recently parameter estimation schemes
 based on Bayesian
statistics using  Markov chain Monte-Carlo (MCMC) method
also has been proposed and implemented in the ground based detector
context~\cite{ChristMeyer01,RMC06} and the LISA
case~\cite{CornishPorter06,WickStroVecc06}.
Using 2PN waveform
Ref~\cite{CornishPorter06} found that posterior parameter estimation
distribution of the {\it extrinsic parameters} obtained using MCMC methods
are in excellent agreement with those computed using Fisher matrix
whereas there is a systematic overestimate by Fisher matrix for the
{\it intrinsic parameters}.

In the present work we extend using the covariance matrix formalism, the parameter estimation problem 
for supermassive BH binary inspirals in LISA  and study the implications
of higher PN order terms. Coalescences of BHs of masses
$10^4-10^7\,M_\odot$ at a luminosity distance of 3 Gpc are considered.
 We assume LISA will observe these events for one year duration.
Using the 3.5PN phasing we calculate the errors associated with the
estimation of the  mass parameters and coalescence time
and
compare them with the corresponding 2PN results.
We also study, for chosen source locations and orientations, 
 the effect of orbital
motion for parameter estimation by comparing these results with
the other two cases: one where LISA is assumed to be a single
interferometer and
another when LISA is
considered to be a two detector network. Conclusions drawn from the
exercise above is far from general and have to be supplemented in future
by rigorous Monte Carlo simulations.

The rest of the paper is organized as follows. Sec.~\ref{sec:PELisa}
discusses all the necessary inputs required for the paper such as
a brief introduction to parameter estimation using covariance matrix,
noise model for LISA, model for the waveform and some other conventions
followed in the paper. Secs~\ref{sec:NoPat} and \ref{sec:WithPat} discuss the main results
and their implications and Sec.~\ref{sec:summary} provides the summary
and future directions.

\section{Parameter estimation for LISA with 3.5PN
phasing}\label{sec:PELisa}
\subsection{Parameter estimation using the covariance matrix}\label{sec:PE}
We summarize the theory of parameter estimation in the context of
Gaussian random detector noise, addressed in the GW context first by Finn and
Chernoff~\cite{Finn92,FCh93} and implemented in detail by Cutler and Flanagan~\cite{CF94}.
Let us assume an inspiral GW signal is detected meeting the necessary
detection criteria and one needs to extract the intrinsic and extrinsic
parameters from the signal by matched filtering.

For sources like inspiralling compact binaries, where a prior source 
modelling is possible to predict the gravitational waveform, {\it
matched 
filtering} is an ideal method both for detection as well as parameter 
estimation of the signal \cite{Wainstein}.  In matched filtering, the 
detector output is filtered using a bank of theoretical templates with 
different signal parameters. The parameters of the template which 
obtains the best signal-to-noise ratio (SNR) gives the ``measured''
values 
of the signal parameters.  These values, in general, will be different 
from the ``actual'' values due to the presence of noise. The problem 
of parameter estimation addresses the question of how close are the 
measured values to the actual ones and what the associated 
errors are in the estimation of different parameters. For a given
signal, different realizations 
of noise lead to different sets of best-fit parameters of the signal.  
When the background noise is a stationary, random, Gaussian process, 
at high enough SNRs the best-fit values of the parameters have a 
Gaussian distribution centered around the actual values of the
parameters.

If ${\lambda^i}$ denotes the actual value of the parameters
and ${\lambda^i}$+$\Delta {\lambda^i}$, the measured value, 
then the root mean square difference $\Delta {\lambda^i}$ obeys a Gaussian distribution:
$p(\Delta{\lambda^i})\propto \exp\left
(-\Gamma_{ij}\Delta\lambda^i\Delta
\lambda^j\right/2 )$ where $\Gamma_{ij}$, the Fisher information
matrix constructed from the Fourier domain representation of the
waveform, is 
given by
\begin{equation}
\Gamma_{ij} = 2 \int_{f_{\rm in}}^{f_{\rm fin}}
\frac{\tilde{h}_{i}^*(f) \tilde{h}_{j}(f) +
\tilde{h}_{i}(f) \tilde{h}_{j}^*(f)}{S_h(f)}\; df.\label{eq:gamma-eqn}
\end{equation}
Here, $\tilde{h}_{i}(f):=\partial\tilde{h}(f)/\partial{\lambda^i}$, 
$\tilde{h}(f)$ is the Fourier domain gravitational
waveform  and $S_h(f)$ is the  (one-sided) noise power spectral 
density of the detector. $f_{\rm in}$ and $f_{\rm fin}$
denote the lower and upper limits of integration which is defined
later in the section.
It also follows that the root-mean-square errors are given by
$\sigma^i=\sqrt{\Sigma^{ii}}\label{eq:sigma_i},$ where
$\Sigma=\Gamma^{-1}$ is called the {\it covariance matrix}.
The non-diagonal elements of the covariance matrix are related to
the correlation coefficient between two parameters $\lambda^i$
and $\lambda^j$:
$c^{ij}:=\frac{\Sigma^{ij}}{\sqrt{\Sigma^{ii}\Sigma^{jj}}}$.
Repeated indices are {\it not} summed over in the above expressions.
Finally the SNR can be expressed in terms of the Fourier
domain signal $\tilde{h}(f)$ as
\begin{equation}
\rho^2 = 4 \int_{f_{\rm in}}^{f_{\rm fin}}
\frac{|\tilde{h}(f)|^2}{S_h(f)}\; df.\label{eq:SNR}
\end{equation} 
In the present case the $\lambda^i$ denoting our chosen set of parameters
are given by $\{t_c, \phi_c, {\cal M}, \eta, D_L, \bar{\phi}_S,
\bar{\phi}_L,
\bar{\theta}_S, \bar{\theta}_L\}$. In the case where pattern functions are not
included the above set reduces to $\{t_c, \phi_c, {\cal M}, \eta\}$. The additional elements of the parameter set
denotes the distance, orientations and locations of the source in the
sky specified with respect to the fixed solar system based coordinate system.

In the above integrals, the upper  limit of integration is
$f_{\rm fin}={\rm Min}[f_{\rm lso},f_{\rm end}]$, where
$f_{\rm lso}$ is the frequency of the innermost stable circular orbit
for the test particle case, $f_{\rm
lso}=({6^{3/2}\,\pi\,m})^{-1}$ and  $f_{\rm upper}$
corresponds to the upper cut-off of the LISA noise curve $f_{\rm
end}=1{\rm Hz}$.
We have chosen the lower limit of frequency $f_{\rm in}={\rm Max}[f_{\rm
in}, f_{\rm lower}]$ where $f_{\rm in}$ is calculated
 by assuming the signal to last
for one
year in the LISA sensitivity band and $f_{\rm lower}$, the low frequency
cut-off for LISA  noise curve, is assumed to be $10^{-5}{\rm
Hz}$\footnote{Another way of choosing the limits of integration
is to calculate the time over which the signal will last once it enters
the LISA band. See \cite{Hughes02} for example, where the duration of the
signal is calculated using the expression for $t(f)$. It assumes $f_{\rm
lower}=10^{-4}$Hz and  3 year mission time for LISA.}.
This is a rather optimistic choice for the lower cut-off for
LISA and in Sec.~\ref{sec:cut-off} we estimate its effect on parameter
estimation 
by comparing it to the more modest choice of $10^{-4}$ Hz.

We follow the noise model of LISA as given in Sec. IIC of Ref~\cite{BBW05a} 
which is a slightly modified version of~\cite{BarackCutler04}.
The noise spectral density consists of a non-sky averaged
part~\cite{BarackCutler04} and confusion noise due to the galactic and
extra galactic white dwarf binaries~\cite{NYP01,FarmerPhinney03,BarackCutler04}.
The explicit expressions for these components can be found in Eqs (2.28)-(2.32)
of Ref~\cite{BBW05a}.
\subsection{LISA detector configuration and the waveform model}
LISA is a three-arm interferometer where each arm has a length of
$5\times10^6\,{\rm km}$. These arms form an equilateral triangle and move in a heliocentric orbit
with a 20$^\circ$ lag to the earth and the plane of the detector tilted at
60$^\circ$ with respect to the ecliptic (See \cite{Bender95,Cutler98} for details). LISA with three arms is essentially equivalent to a pair of
two-arm detectors capable of simultaneously measuring the two polarizations
of the incoming gravitational wave.
We consider two cases: one where we assume the
estimation of mass parameters is not affected because of their
correlations with the angular variables and second when we estimate the
associated errors with angular variables and luminosity distance of the
source. Since the information about the angular variables are encoded in
the so-called pattern functions which describes the orbital motion of
LISA, in the first case we use a waveform which is averaged over the
pattern functions. In the second case, we do not average over the
pattern functions and use the information from the LISA orbital motion
to discuss the estimation of angular resolution and luminosity distance
to the
source. Further, in the second case we consider  cases when (i)
LISA is a single two arm-detector and (ii) as a two detector network
in order to understand the effect of network configuration for parameter
estimation.
The LISA antenna patterns, describing its orbital
motion, is given in ~\cite{Cutler98} which is used for the present
study (also see Appendix A of Ref~\cite{BBW05a} for these expressions). 

Unlike the ground based detectors where the two arms have an angle
90$^\circ$, the LISA arms are at 60$^\circ$. As shown by
Cutler~\cite{Cutler98}, the relative strain amplitude (which is 
the gravitational waveform) in the LISA case can be simply be related to
the 90$^\circ$ interferometer case by multiplying the latter by a factor
${\sqrt 3}/2$. Using this input, the Fourier domain waveform within the
stationary phase approximation can be written down
as~\cite{Cutler98,BBW05a} 
\begin{subequations}\label{eq:NPWF}
\begin{eqnarray}
\tilde h_\alpha(f) &=&\frac{\sqrt{3}}{2}\,{\cal A}\,f^{-7/6}\,e^{\rm{i}
\psi(f)}\,, 
\quad \quad \alpha = {\rm I,II}\,,
\\{\cal A} &=& \frac{1}{\sqrt{30}\pi^{2/3}} \frac{{\cal M}^{5/6}}{D_{\rm
L}}\,,
\end{eqnarray}
\end{subequations}
where $\alpha$  labels the interferometer, $f$ the GW
frequency and ${\cal M}$, the chirp mass which is related to total
mass $m=m_1+m_2$ and symmetric mass ratio $\eta=m_1m_2/m^2$ by
${\cal M}=\eta^{3/5}\,m$. The luminosity distance to the source is denoted
by $D_L$. The GW phase $\psi(f)$ appearing in the formula is completed
up to 3.5PN~\cite{BDEI04,BFIJ02,BIJ02} and its Fourier domain
representation is given in ~\cite{DIS01,DIS02}. We find it more
convenient to write it as
\begin{equation}
\psi(f)=2\pi\,f\,t_c-\phi_c+\sum_{k=0}^{k=7}\alpha_k\,v^k\;,
\end{equation}
where $v=(\pi\,m\,f)^{1/3}$ is the PN variable which is related to 
gauge independent source velocity in system of units where $G=1=c$ which
we follow henceforth in the paper. Eq~(3.4) of \cite{AISS05} gives the 
 $\alpha_k$ for different values of $k=0\cdots 7$.

In the case where we do not average the pattern functions, the waveform
can be written as~\cite{BBW05a} 
\begin{eqnarray}
\label{WPWF}
\tilde h_\alpha(f) = \frac{\sqrt{3}}{2}
{\cal A}\, f^{-7/6}\,e^{i \Psi(f)}\, \left \{ \frac{5}{4}
\tilde{A}_\alpha(t(f)) \right \}
e^{-i\bigl( \varphi_{p,\alpha}(t(f)) +  \varphi_D(t(f))\bigr)}
\,,
\end{eqnarray}
where $\varphi_{p,\alpha}(t(f))$ and $\varphi_{D}(t(f)$ are the
polarization phase and Doppler phase respectively~\cite{Cutler98}.
$\tilde{A}_\alpha(t(f))$ correspond to
the amplitude modulations induced by the LISA's orbital motion.
$\tilde{A}_\alpha(t(f))$ and $\varphi_{p,\alpha}(t(f))$ depends on
the pattern functions $F^{\alpha}_{+}(t)$ and $F^{\alpha}_{\times}(t)$ and
hence vary with time. For the
explicit expressions for  $\varphi_{p,\alpha}$ and $\varphi_{D}$,
we refer the readers to Refs~\cite{Cutler98,BBW05a}. For 3.5PN accurate
expression for $t(f)$ we use the following relation
\begin{eqnarray}
2\pi\,t(f)&=&  \frac{d\,\psi(f)}{df}.
\end{eqnarray}
This can be rewritten as
\begin{eqnarray}
t(f)&=& t_c-\sum_{k=0}^7{t^v_k\,v^k},
\end{eqnarray}
and values of $t^v_k$ is given in Refs~\cite{DIS01,DIS02} which can
readily be used.

For calculations where LISA is assumed to be a two detector network,
we calculate the SNR and Fisher matrix using
\begin{eqnarray}
\rho^{\rm Network}=\sqrt{\rho_{\rm I}^2+ \rho_{\rm II}^2},\\
\Gamma_{ab}^{\rm Network}=\Gamma_{ab}^{\rm I}+ \Gamma_{ab}^{\rm II}.
\end{eqnarray}
The errors for the two detector case are obtained inverting the total
Fisher matrix following the procedure outlined in Sec.~\ref{sec:PE}.

Throughout the paper we assume a cosmological model with zero spatial curvature
$\Omega_{\kappa}=0$, $\Omega_{\Lambda}+\Omega_{M}=1$ and Hubble's
constant to be $H_0=70$ km s$^{-1}$ Mpc$^{-1}$. $\Omega_{\Lambda}$ and
$\Omega_{M}$ refers to the contributions to the total density from
matter and cosmological constant.
The luminosity distance is given by
\begin{equation}
D_L=\frac{1+z}{H_0}\int_0^z
\frac{dz'}{\left[\Omega_M(1+z')^3+\Omega_\Lambda\right]^{1/2}}\,,
\end{equation}
where $z$ denotes the redshift of the source.

We calculate the Fisher matrix for the different configurations
of LISA using the corresponding waveforms and invert it to get the
covariance matrix. The elements of the covariance matrix
are used for discussing the errors and correlation coefficients
of different parameters in the next section.
While discussing the trends with the PN orders, its useful to keep
in mind that
 when pattern functions are included,
there are
additional PN expansions coming from $t(f)$ both in the amplitude and
phase, apart from the usual phase $\psi(f)$, which can influence the
results.
\section{Parameter estimation with pattern averaged
waveform}\label{sec:NoPat}
In this section we discuss the parameter estimation in the LISA
case using pattern averaged waveform of Eq~(\ref{eq:NPWF}).
Parameter estimation with the non-pattern averaged waveform 
(as in Eq~\ref{WPWF}) is more complex
since the estimated errors strongly depend on the {\it location and
orientation}
of the source (see the Sec.~\ref{sec:WithPat} below) which enter the
calculation via
the pattern functions. The ideal way to deal with the
situation will be to perform Monte Carlo simulations for different
binaries located and oriented randomly in the
sky~\cite{Hughes02,Vecchio04,BBW05a,BBW05b}.
A recent Monte Carlo simulation~\cite{BBW05a}, which addressed
the parameter estimation problem using the 2PN waveform including spin
effects, compared the results of their simulation with the result
they obtained using pattern averaged waveform (see Tables V, VI and VIII
of Ref~\cite{BBW05a}). They found that the results in both
the cases are in excellent agreement.
The results presented in this section about the improved parameter
estimation with the pattern averaged waveform, may hence give
a reasonably good idea about the full problem where the
LISA pattern functions are included and it is considered to be a two
detector network.
We emphasize that the results quoted here have to be supported by
 Monte Carlo
simulations similar to Ref~\cite{BBW05a} (see concluding remarks in
Sec.~\ref{sec:summary}). 

\subsection{Improved parameter estimation of equal mass binaries with the 3.5PN phasing}
We discuss the performance of the 3.5PN restricted waveform from the parameter estimation
point of view for the pattern-averaged case discussed above.
Our aim is to study the variation of errors in different parameters with
the
total observed mass\footnote{By total mass,  we always refer to
the total redshifted mass $m'(1+z)$, where $m'$ is the actual source
mass and  $z$ is the redshift of the
source. This is the mass parameter that is measured by the GW observations.} of
the binary for different PN orders.
This would not only give us an idea of the improvement
brought in by the use of higher order phasing but also about the convergence of the PN series for the problem of
parameter estimation. 
We have checked our codes by 
reproducing the results
at 2PN with that in Table III and V of Ref~\cite{BBW05a} for the nonspinning
case.
The important results of our study are
discussed in detail in what follows. The errors in estimation
of different parameters for a $2\times10^6M_\odot$ binary at 3 Gpc is
provided in Table~I for different PN orders in the phasing.

\begin{figure*}
\centering
\includegraphics[height=2.0in,width=2.5in]{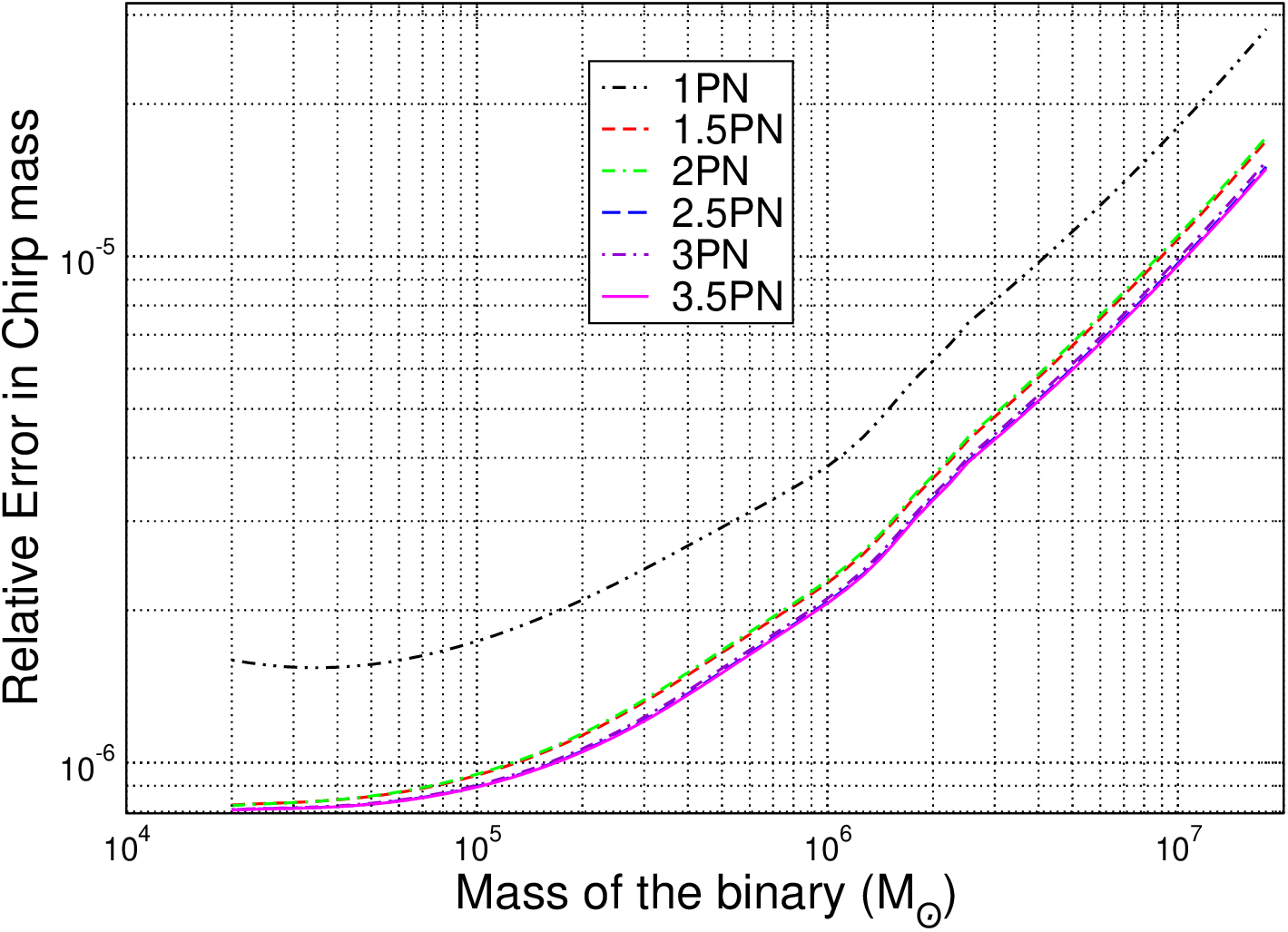}
\hskip 0.1 true cm
\includegraphics[height=2.0in,width=2.5in]{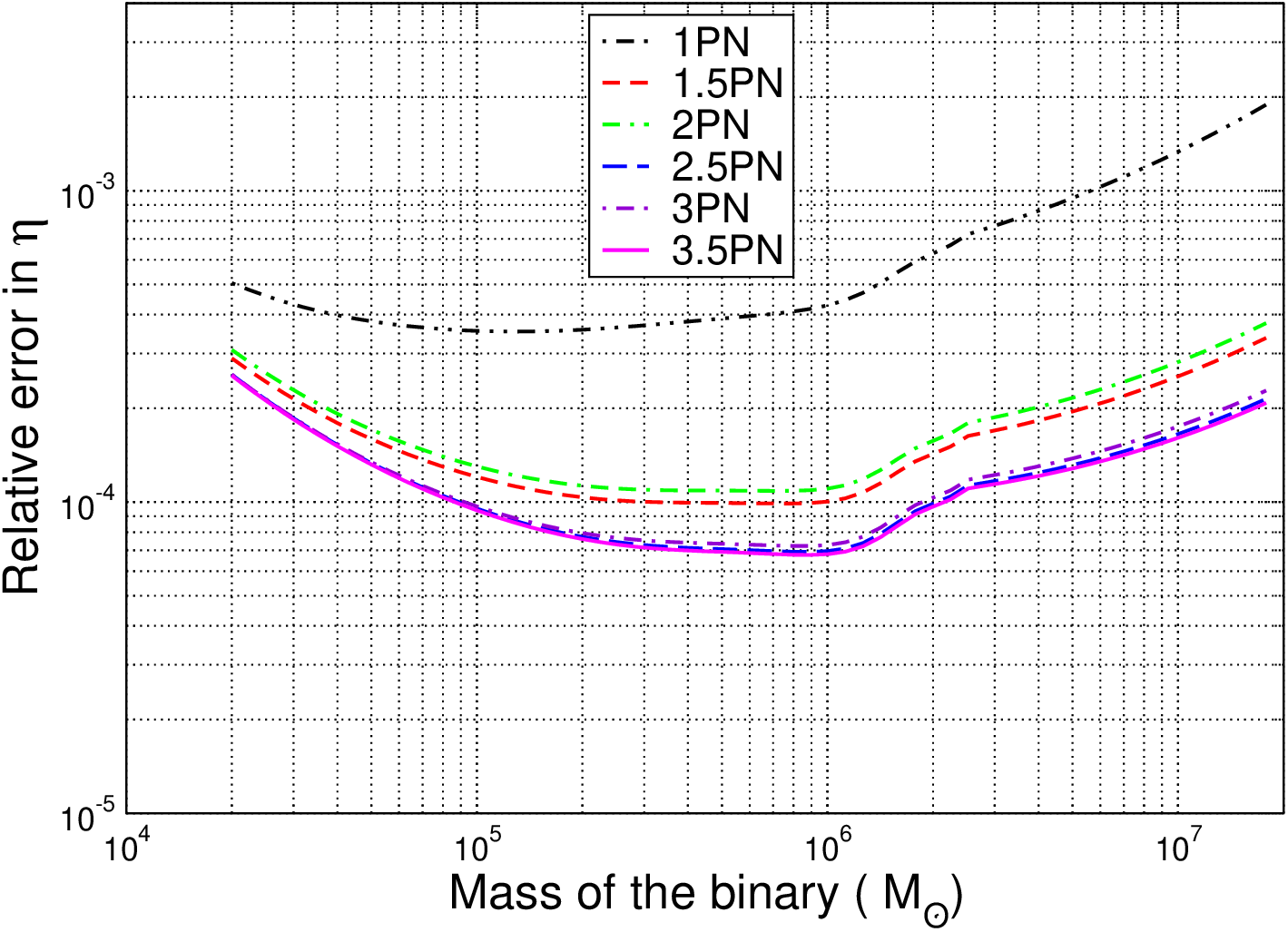}
\caption{Variation of errors with the total observed mass for different PN
order restricted
waveforms for LISA. {\it Pattern averaged} waveform is used.
The convergence of the results is evident in both the
cases. Sources are assumed to be at 3 Gpc. 
}
\label{Fig:NoPattern}
\end{figure*}

\subsection{Improvement in estimation of mass parameters and PN
convergence}
{\it Improvement due to higher order terms}: We plot in Fig.~\ref{Fig:NoPattern} the variation with mass
of the errors in chirp mass and $\eta$ for different PN orders.
There is significant improvement in the estimation by the use of the
 3.5PN phasing
instead of the 2PN one especially for more massive systems. For a
prototypical system of a binary BH each of
mass $10^6 M_{\odot}$, we find that the chirp mass and $\eta$
improve by $11\%$ and $39\%$ respectively. Improvement is higher for
more massive systems. For a $2\times10^7M_\odot$ binary, the chirp mass
and $\eta$ improves by $14\%$ and $45\%$. They are similar to the
results for the ground based detectors as discussed in ~\cite{AISS05} but at
an entirely different mass range.
For a typical binary in its sensitivity band, LISA will be able to measure chirp mass with an
incredibly small
 fractional accuracy of $\sim 10^{-6}$ and $\eta$ by about
$\sim10^{-4}$.

{\it Variation with mass:} The estimation of the chirp mass worsens with
increase in total mass of the binary whereas the estimation of $\eta$
improves initially and then decreases. These effects can be understood
as follows. When the total mass increases there are two competing
effects in action: the increase in errors with mass, since signal lasts
for smaller duration, 
and the variation of SNR with mass, which is a 
characteristic of the noise curve. For chirp mass the errors increase so rapidly
that the variation in SNR does not affect the trend and the errors
continue to increase monotonically with
mass. For $\eta$, there is trade-off between these two competing effects
which accounts for the minimum in the curve.
\subsection{Errors in coalescence time}\label{sec:tc}
Measuring the time of coalescence of a binary system is important
to carry out electro-magnetic observation of the
event associated with the binary merger. We discuss the trends in
 estimation of $t_c$ below.

\begin{figure*}[t]
\centering
\includegraphics[height=2.5in,width=2.5in]{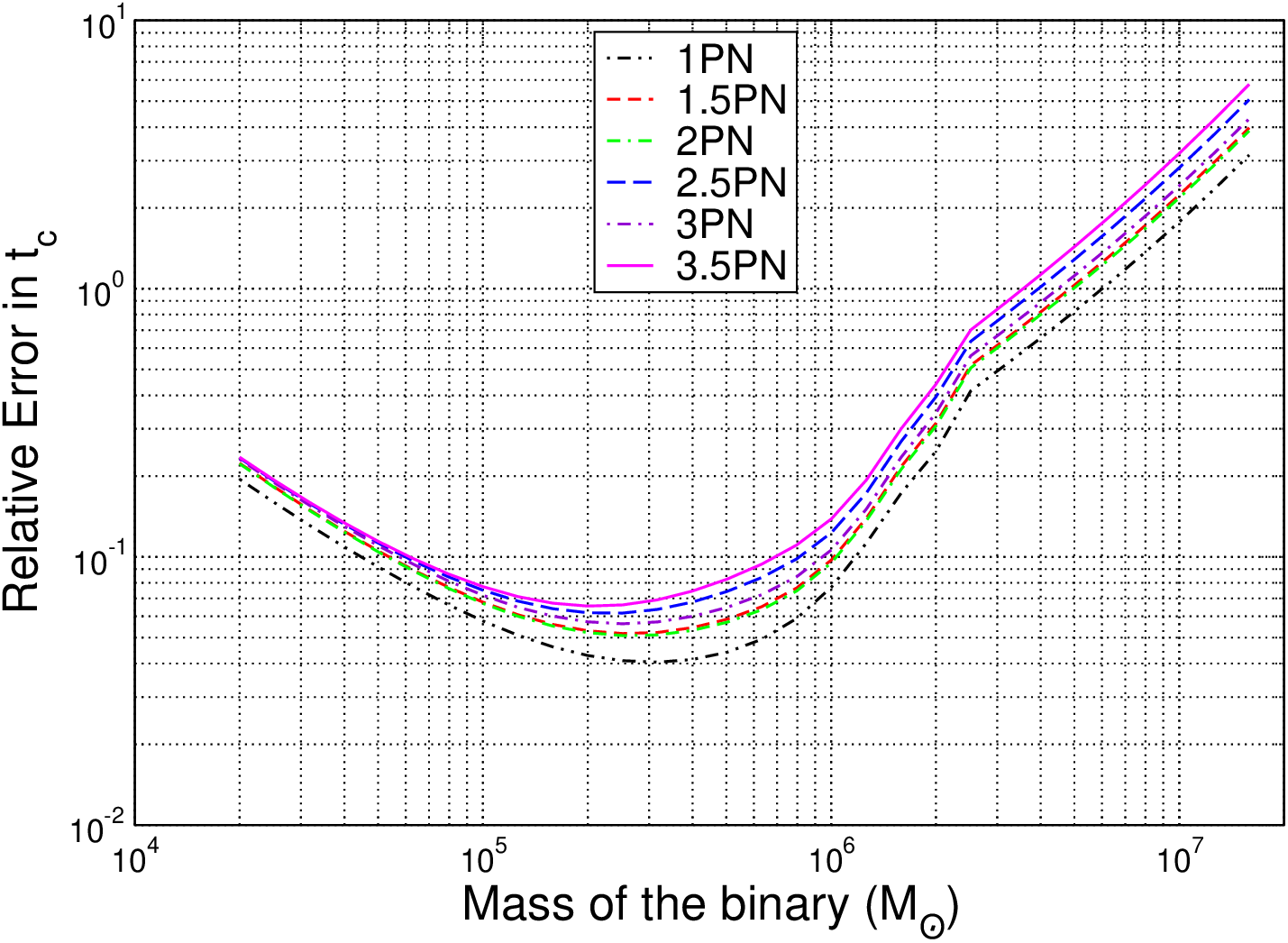}
\caption{Variation of errors in $t_c$ 
with the total observed mass for different PN
order restricted
waveforms for LISA.  {\it Pattern averaged} waveform is used.
The convergence of the results is evident from the plot.
 Sources are assumed to be at 3 Gpc. 
}
\label{Fig:NoPatterntc}
\end{figure*}

Fig.~\ref{Fig:NoPatterntc} displays the variation of errors in $t_c$
with increasing mass of the source and across the PN orders.
 The errors in $t_c$ show trends
similar to that of~\cite{AISS05}, {\it i.e.,} with increase in PN order
the errors oscillate in a sense opposite to the mass parameters and in going from
2PN to 3.5PN there is a net degradation in its estimation, which is about
43\% for the $2\times10^{6}M_\odot$ system considered.
This was explained in Ref~\cite{AISS05} based on
the correlations between $t_c$, ${\cal M}$ and $\eta$. It was noticed
that {\it both} $c_{{\cal M}t_{c}}$ and $c_{\eta t_c}$ are positive 
and  follow the same trend as the error in $t_c$. Increase in these correlations implied
a worsened estimation of $t_c$.

\subsection{Post-Newtonian convergence in the parameter estimation
context}
Since the PN series is an asymptotic series, the rate of convergence of the
results is a very
important issue for detection as well as parameter estimation.
We use the word `convergence' 
to mean that the difference (in errors) between two consecutive PN
orders is smaller as we go to higher orders considered.
As remarked in Refs.~\cite{CF94,PW95}, if the parameter estimation
scheme is based
on a lower order (2PN)
phasing, the systematic errors due to the absence of higher order terms
may be more than the statistical errors caused by the noise.
Since we study here the implications of 3.5PN phasing, we examine the
convergence of the series based on our results for different PN orders.

As Fig.~\ref{Fig:NoPattern} reveals, though there will
be improvement by using the 3.5PN phasing instead of the 2PN one, most of the
improvement seems to come from the transition between 2PN and 2.5PN after
which
the series continues to show its characteristic oscillatory behaviour,
but with  smaller magnitude, suggesting that phasing at orders higher than 3.5PN
may not cause much improvement (see Table~I). 
\begin{table}[t]
\centering
\caption{Variation of errors in different parameters and number of
GW cycles with PN order. Errors are calculated with the pattern averaged
waveform. The
system considered is a binary of mass $2\times10^6 M_{\odot}$ at a
luminosity distance of 3 Gpc. Most of the features regarding the
improvement in parameter estimation, convergence of the PN series and
correlation with number of GW cycles are captured in this table.}
\vskip 12pt
\begin{tabular}{@{}lccccccccccccccc@{}}
\hline
\hline
PN Order
&\vline&$\Delta t_c$ &\vline&$\Delta {\cal M}/{\cal M}$
&\vline&$\Delta \eta/\eta$
&\vline&$N_{\rm cycles}$\\
&\vline& (sec)&\vline&($10^{-6}$) &\vline& ($10^{-4}$)&\vline&
\\
\hline
\hline						   	     		     
1PN &\vline& 0.2474  &\vline&6.217
&\vline&6.287 &\vline&2414.03\\
1.5PN &\vline&0.3149   &\vline&3.648
&\vline&1.427 &\vline&2310.26\\
2PN &\vline&0.3074&\vline&3.694
&\vline&1.572 &\vline&2305.52\\
2.5PN &\vline& 0.3947&\vline& 3.320  &\vline&0.9882
&\vline&2314.48\\
3PN &\vline&0.3435 &\vline&3.377   &\vline&1.033&\vline&2308.73\\
3.5PN &\vline&0.4399 &\vline&3.300&\vline&0.9661&\vline&2308.13\\
\hline
\hline
\end{tabular}
\end{table}

\subsection{Parameter estimation and Number of GW cycles}
In Ref.~\cite{AISS05}, the correlation between the improvement in errors
across different PN orders and the number of total and useful GW
cycles~\cite{DIS00} was studied. It was found that though they are good indicators
of how the errors at each order vary with the total mass of the system, 
they alone cannot explain the variation of errors across different PN
orders in the context of ground based detectors.
We confirm this feature in the LISA context. We recover the results in
Table I and II of Ref~\cite{BBW05a} as a check of our calculation.
Table~I shows how the errors and total GW cycles vary with increasing PN orders. One would expect
an improved (worsened) estimation if number of GW cycles increases
(decreases) between two consecutive PN orders if they were solely
responsible for the trends. 
From the table it is clear that from 1PN to 1.5PN
and 1.5PN to 2PN, the errors in chirp mass and $\eta$ do not conform to the
above expectation.
The same is the case in going from 3PN to 3.5PN. Finally, trends
in $t_c$ being opposite to that of the other two mass parameters lead one to 
conclude once again that 
the number of GW cycles is not sufficient to understand the variation
of errors with PN orders.

Irrespective of whether the total number
of GW cycles is very high ($\sim 10^5$)(as in the case of LISA) or
low ($\sim$ hundreds) (as for the ground based detectors) the PN trends in parameter
estimation are too complicated to be explained solely in terms of this.
However the  variation of errors with mass can be understood
on the basis of number of the cycles and variation in SNR (again, not in terms
of only one of them).

\subsection{Parameter estimation for unequal mass
binaries}\label{sec:Unequal}
Lastly we perform a similar analysis for unequal mass systems where
$\eta <0.25$.
For larger mass ratios ($\eta\simeq10^{-5}$ or smaller in our case), the
Fisher matrix becomes ill-conditioned~\cite{BBW05a} and hence
we restrict ourselves to the inspiral of a binary consisting of an
intermediate mass BH (IMBH)
 and a SMBH  rather than stellar mass-SMBH
inspirals.
An  IMBH of $10^4M_\odot$  inspiralling into a SMBH
of $10^6 M_\odot$ constitutes our prototypical system. This system
at a distance of 3 Gpc will have a SNR of a few hundreds. 
We find that the
improvement due to the inclusion of higher order terms is more dominant here
than for the equal mass binary case. For the prototypical system considered
above, we find an improvement of $11\%$ for the chirp mass and $52\%$
for $\eta$ in the pattern-averaged case. For a $10^4-10^7M_\odot$
binary, where SNR is $\simeq 100$, the improvement is even more: $20\%$
and $62\%$ respectively\footnote{Using  a calculation of the number of
GW cycles, Ref~\cite{Berti06} has
emphasized the  need for higher order PN modelling of the IMBH-SMBH binaries.
The  effects of eccentricity could also play an important role in the dynamics
of such binaries.}.

This larger improvement for the unequal mass binaries is not a special
feature of the LISA noise curve;  for the ground based detectors
also a similar feature exists. 
But unlike in the LISA case, where many such unequal mass binaries are
astrophysically plausible, for the ground based detectors
such sources are not prototypical.

\subsection{Effect of low frequency cut-off chosen}\label{sec:cut-off}
All the calculations so far, and hence the results, have been based on
the optimistic possibility that
the low frequency sensitivity of LISA can be extrapolated from
$10^{-4}$ Hz to $10^{-5}$ Hz. As argued in Ref~\cite{BBW05a}, this
significantly improves the parameter estimation. 
We quantify the effect of this choice of lower cut-off
by comparing our previous results
with the one where LISA is assumed to be `blind' below $10^{-4}$Hz. 

\begin{figure*}[t]
\centering
\includegraphics[width=5in]{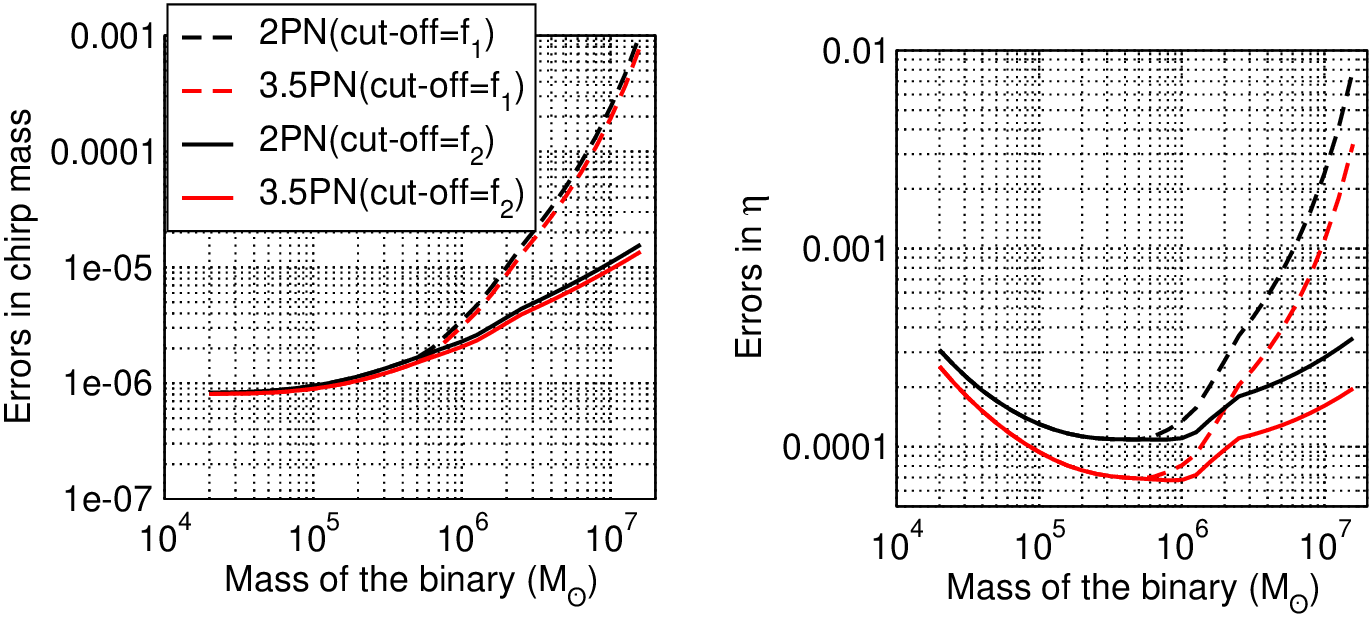}
\caption{Variation of errors of chirp mass
and $\eta$ 
with the total observed mass for different PN
order restricted
waveforms for LISA.  {\it Pattern averaged} waveform is used. for the two
choices of the lower frequency cut-offs $f_1=10^{-4}$ Hz and
$f_2=10^{-5}$ Hz.
Sources are assumed to be at 3 Gpc.}
\label{Fig:Cutoff}
\end{figure*}

As discussed in Sec.~\ref{sec:PE}, we have chosen the lower limit 
of integration in all the calculations assuming that the system is
observed for one year before coalescence, when $f=f_{\rm
lso}$. By this procedure, the lower cut-off for a
$2\times10^{5.5}M_\odot$ binary is about $10^{-4}$ Hz. This means 
for systems with masses higher than $2\times10^{5.5}M_\odot$ (and hence
a lower $f_{\rm lso}$) the lower limit of integration for one year observation
time will be less than $10^{-4}$ Hz. If we assume LISA is not sensitive
to signals below $10^{-4}$ Hz, these systems 
 will be observed effectively for less than a year
leading  to a significant decrease in the number of GW cycles and
consequent degradation in
parameter estimation (for a $2\times10^6 M_\odot$ binary, a lower cut-off
of $10^{-5}$ Hz will give 2308 GW cycles as opposed to 608 if the
cut-off was $10^{-4}$ Hz).
Thus for binaries whose masses
are higher than $2\times10^{5.5}M_\odot$, the choice of lower cut-off frequency will affect
the results reported here. Fig.~\ref{Fig:Cutoff} displays the variation of
2PN and 3.5PN errors in chirp mass and $\eta$ corresponding to the
two different lower frequency cut-offs we have chosen. Indeed, as is evident
from the plot, 
the errors start to deviate for binaries whose masses are greater than
 $2\times10^{5.5}M_\odot$.
For a $2\times 10^7M_\odot$ system, using
$10^{-5}$ Hz as cut-off instead of $10^{-4}$ Hz
improves the estimation of chirp mass  by about 150 times and that
of $\eta$  by 40 times.
These results  confirm the need to
push to the extent possible the lower frequency sensitivity of
LISA.

Regarding the improvement in parameter estimation in going from 2PN to
3.5PN, calculation with a cut-off of $10^{-4}$ Hz shows that for a
$2\times10^7M_\odot$ binary the difference in going from 2PN to 3.5PN
would be $22\%$ and $60\%$ (as opposed to $13\%$ and $45\%$ with
$10^{-5}$ Hz) for chirp mass and $\eta$. 
 The number of GW cycles for a cut-off of $10^{-4}$ Hz
is just 7 whereas with a $10^{-5}$ Hz cut-off it is about 540.
Therefore with a cut-off of $10^{-4}$ Hz, one is only observing the very late
inspiral of the system\footnote{Inspiral waveforms would be inadequate in
this studies. Theoretical approximants, e.g. Effective
one body ~\cite{BuonD98,BuonD00} may have to be employed to
model this phase of the binary's dynamics.} whereas with a cut-off of $10^{-5}$ Hz, the
inspiral phase is dominant. The significantly larger variation in
parameter estimation in going from 2PN to 3.5PN with the $10^{-4}$ Hz
cut-off could be due to the generally accepted fact that higher order
terms in the phasing formula are more important as one approaches the
last stable orbit.

\section{Parameter estimation without pattern
averaging}\label{sec:WithPat}
Having discussed in detail the various aspects of parameter estimation
using pattern averaged waveform and factors which affect the process,
we now turn our attention to the parameter estimation without
pattern averaging. Obviously the source's location, orientation
and luminosity distance to the source gets added to the space of parameters
which was four dimensional in the previous case. Unlike the ground based detectors, LISA can measure the distance, location and orientation
of the source with
a {\it single} detector because of the modulations due to its orbital motion~\cite{Bender95,Cutler98}. Besides, using LISA
as a two detector network improves the estimation of angular
resolution of the source~\cite{Cutler98}. In this section
we will discuss the improvement brought in by the higher order
terms using non-pattern averaged waveform for LISA.
We check our code, which now includes the pattern functions,
by reproducing the results of~\cite{Cutler98} at 1.5PN with their
signal and noise models.

But as mentioned earlier the strong dependence, as we shall
discuss, of the errors on the angular variables makes our analysis, which 
is for selected values of them following Cutler~\cite{Cutler98}, less
rigorous. The best way to deal with this situation is to
perform Monte Carlo simulations, similar to \cite{Hughes02,Vecchio04,BBW05a,BBW05b}.
However we notice that certain general conclusions can still be drawn
from our limited analysis which is the topic of discussion of this section. 

\begin{figure}[t]
\begin{center}
\includegraphics[width=6in]{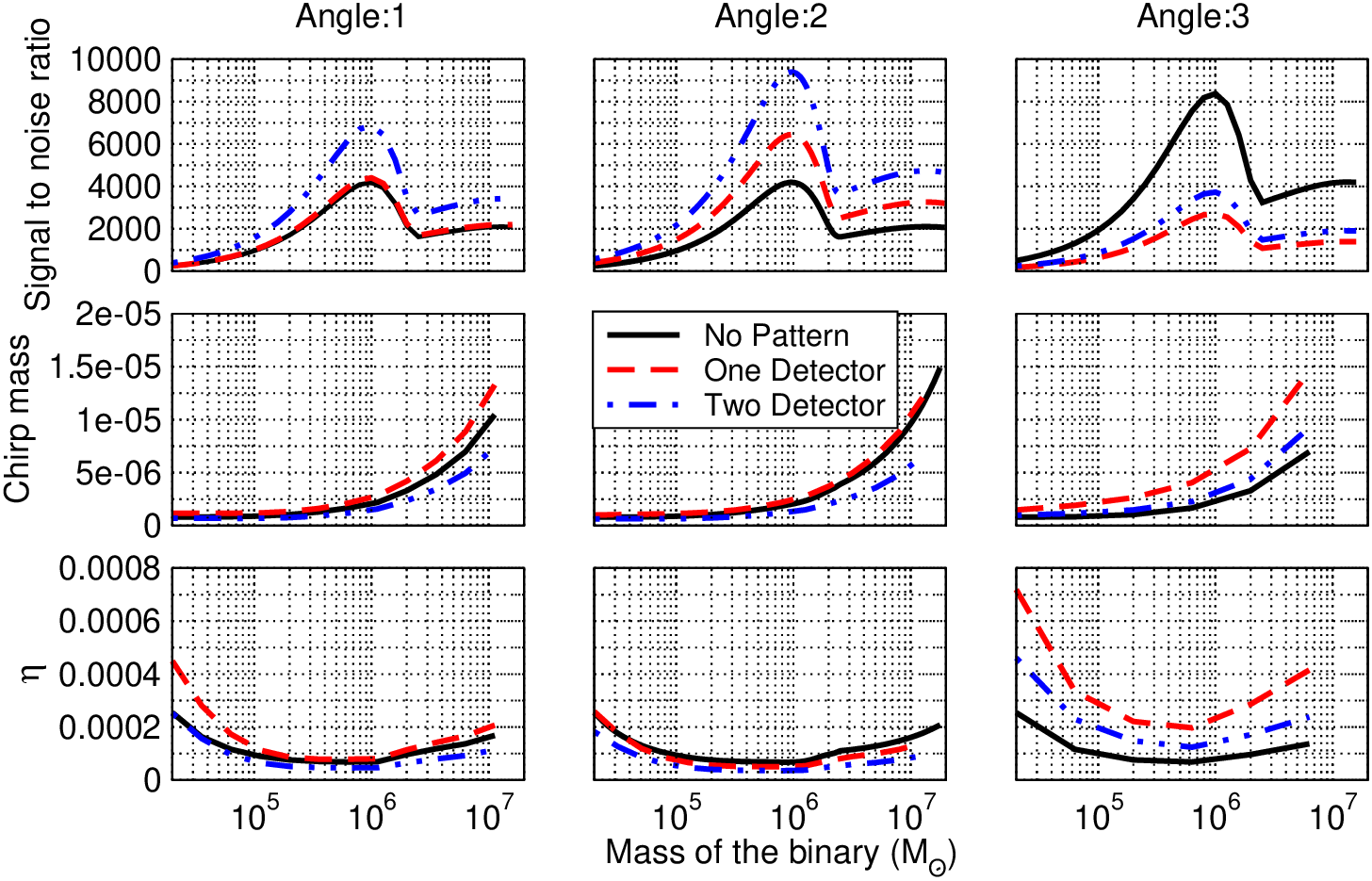}
\end{center}
\caption{Variation of the signal to noise ratio, relative errors in
chirp mass and relative errors in $\eta$ with the total observed mass of the binary for different choices of location
and orientation of the source.
Sources are assumed to be at a luminosity distance of 3 Gpc and a
{\it non-pattern averaged} waveform is used. Angle:1 corresponds to $\{\bar{\mu}_{
L}=0.5,\bar{\mu}_{S}=-0.8,\bar{\phi}_{L}=3,\bar{\phi}_{ S}=1\}$. Angle:2
is $\{\bar{\mu}_{
L}=0.2,\bar{\mu}_{S}=-0.6, \bar{\phi}_{L}=3,\bar{\phi}_{ S}=1\}$ and
Angle:3 $\{\bar{\mu}_{
L}=0.8,\bar{\mu}_{S}=0.3, \bar{\phi}_{L}=2,\bar{\phi}_{ S}=5\}$.
The errors thus depend very much on the
position and orientation of the source in the sky.}
\label{Fig:DiffConfig}
\end{figure}

\subsection{Comparison of different detector configurations}
We compare the estimation of errors with different detector
configurations now.
The final
errors, in comparison with the pattern averaged case, will depend on (i)
the value of SNR corresponding to 
the set of angles chosen (ii) The change in SNR relative to
the pattern averaged case and (iii) worsening of errors due to
the introduction of the new parameters. 
 
For the pattern averaged case, since there are no pattern dependent parameters, the total
parameter space is essentially 4 dimensional, the parameters being
$\{t_c, \phi_c, \cal M,
\eta\}$. When we make a transition from the pattern averaged to the
non-pattern averaged case, five new parameters $\{ D_{\rm L}, \bar{\mu}_{ L},\bar{\mu}_{S},\bar{\phi}_{L},\bar{\phi}_{ S}\}$
corresponding to the distance to the source, its location and
orientation are added to the space of parameters. The parameter space
is now nine dimensional significantly higher than the earlier
four dimensional one.
This increase in dimensionality of the parameter space leads to an increase
in the errors of the four existing parameters. On the other hand, the introduction
of pattern functions results in a change in SNR depending on the four
angles chosen. The final picture is a complex interplay of all these features.

The errors in chirp mass and $\eta$ together with SNR is displayed
in Fig.~\ref{Fig:DiffConfig} for three set of angles from the seven given in ~\cite{Cutler98}.
The three configurations corresponding to the pattern averaged, non-pattern averaged with one detector and the two detector network (without pattern averaging)
are considered.
As  is evident from the plot, the SNR and hence the errors crucially depend
on the location and orientation of the source. Also there can be orientations
which may have lower SNR than for the pattern averaged case
(see third column e.g.).
However an interesting point from the figure (and from those runs corresponding to
other values of angles which are not displayed) is that among the three effects
which influence the parameter estimation without pattern averaging, the value
of SNR seems to be the dominant one. Smallest errors in the plot
correspond to the configuration with the largest SNR. Between the pattern averaged 
and the one detector case, the effect of correlations due to the
additional parameters
play a significant role. For example, in the second column, though one detector case has 
larger SNR, the errors are smaller for the pattern averaged case for this
reason. We do not discuss the percentage improvement arising from the
higher order PN terms since it depends very much on the orientation
of the source. Exhaustive Monte Carlo simulations have to be performed
to have a detailed understanding of this.

We conclude with a remark about the estimation of angular resolution
and distance as observed from the limited set of angles we have considered.
The estimation of distance and angular resolution is not improved much
because of the additional phasing terms. This is not surprising, since
the additional terms in the phasing formula do not carry any information about location
or orientation of the binary.
One may need to go beyond the restricted
waveform model of the waveform in order to achieve this.
 Some preliminary studies in this regard
~\cite{SinVecc00a,SinVecc00b,MH02,HM03,Chris06,ChrisAnand06} are
consistent with the same.
Going beyond the restricted waveform approximation would mean including
the amplitude corrections to the waveform from the two GW polarizations,
currently completed up to 2.5PN order~\cite{BIWW96,ABIQ04}. This is because
the amplitude terms are functions also of the angular positions of the
source in the sky, the introduction of which could break different degeneracies,
allowing better parameter estimation~\cite{MH02,HM03}.

\section{Summary and Future works}\label{sec:summary}
The significance of higher order phasing terms is investigated in the
LISA case for different sources using a pattern averaged waveform model.
Using the 3.5PN inspiral
waveform instead of the 2PN one which is currently employed in the  GW experiments, mass parameters can be estimated with
improved precision for LISA.
For an equal mass binary of
$2\times10^6M_\odot$ at a
luminosity distance of 3 Gpc, the improvement
 in chirp mass is $\sim 11\%$ and that of $\eta$ is $\sim 39\%$.
The PN series shows convergent behaviour beyond 2.5PN order.
The number of GW cycles is a good indicator of how the errors
vary with mass but not across different PN orders. The improvement
in parameter estimation is more pronounced for binaries with unequal
masses. The problem of unequal mass binaries should be revisited with
the inclusion of spin effects to higher orders as spin effects play a
dominant role in the dynamics of such systems. Inclusion of the spin effects at 2.5PN order in the
phase~\cite{FBBu06,BBuF06} would help us go beyond
the results of~\cite{BBW05a,Vecchio04} for the spinning case and will be exciting
especially for the equal mass case.
The effect of the lower cut-off frequency we have chosen ($10^{-5}$ Hz) on the
parameter estimation is studied by comparing the calculation
with the more modest cut-off of $10^{-4}$ Hz.
The estimation of source location and orientation
is also studied using the non-pattern averaged waveform for selected source
directions and orientations.
They do {\it not} improve significantly by the use of the 
restricted 3.5PN template. One may need to go
beyond the restricted waveform approximation in order to achieve it.
Especially, implications of the full waveform with the 2.5PN polarizations
of Ref~\cite{ABIQ04} together with the 3.5PN phasing of \cite{BFIJ02}
would be an interesting exercise to carry out.

Exhaustive Monte Carlo simulations are required in order to understand
the effect of higher PN terms for LISA when the non-pattern averaged
waveform is used since 
the parameter estimation in this case strongly depends on the location
and orientation of the source. We plan to take up this problem in the
near future.

It may be interesting to employ Markov chain Monte Carlo methods
for parameter estimation~\cite{WickStroVecc06,CornishPorter06} and compare the results with the ones obtained
using the covariance matrix.
Extending this to the present context would surely be interesting
and should be addressed.
Finally, the parameter estimation for the SMBH inspirals may eventually
have to be done using the time delay
interferometry variables~(see \cite{TDhu05} and references therein).

The very high accuracy parameter extraction possible with LISA
will make it a useful tool of astrophysics and provide thorough probes of strong field
aspects of gravity in the future.
\acknowledgments
The author thanks B R Iyer for useful discussions, valuable comments
on the manuscript, and constant encouragement.
Conversations with B S Sathyaprakash on issues related the
 parameter estimation with LISA and his encouragement to pursue this
problem are gratefully acknowledged. The author is grateful to
Sanjeev Dhurandhar for discussions and comments on the manuscript.
The author also thanks M S S Qusailah
for useful discussions. I am grateful to the anonymous referee for
valuable comments which have improved the presentation
significantly.

All the calculations reported in this paper are performed with
{\it Mathematica}.
\bibliography{/tphome/arun/ref-list}
\end{document}